\documentstyle[preprint,eqsecnum,aps]{revtex}

\begin{document}
\draft
\preprint{}
\title{charge fluctuations in YBa$_2$Cu$_3$O$_{6+x}$ superconductors \\}
\author{H. A. Mook, $^1$ and F. Do$\rm\breve{g}$an,$^2$}
\address{$^1$Solid State Division, Oak Ridge National Laboratory, Oak Ridge, Tennessee 37831-6393}
\address{$^2$Department of Materials Science and Engineering, University of Washington, Seattle, Washington 98195}
\date{\today}
\maketitle
\begin{abstract}
Striped phases in which spin and charge separate into different regions in the material have been proposed to account for the unusual properties of the high-$T_c$ cuprate superconductors.  The driving force for a striped phase is the charge distribution, which self-organizes itself into linear regions.  In the highest $T_c$ materials such regions are not static but fluctuate in time.  Neutrons, having no charge, can not directly observe these fluctuations but they can be observed indirectly by their effect on the phonons.  Neutron scattering measurements have been made using a specialized technique to study the phonon line shapes in four crystals with oxygen doping levels varying from highly underdoped to optimal doping.  It is shown that fluctuating charge stripes exist over the whole doping range, and become visible below temperatures somewhat higher than the pseudogap temperature. 
\end{abstract}


\narrowtext
Striped phases are inhomogeneous distributions of charge and spin that have been suggested to account for many of the unusual properties of the high-$T_c$ cuprate superconductors \cite{1,2,3,4,5,6,7}.  In the simplest picture the charge and spin can be thought to be confined to separate linear regions in the crystal and thus resemble stripes.  We expect static striped-phases for the YBa$_2$Cu$_3$O$_{6+x}$ materials to not coexist with superconductivity except perhaps, for materials with low-oxygen contents and low values of $T_c$.  However, neutron scattering measurements have shown results for both spin- \cite{8,9} and charge- fluctuations \cite{10} that suggest that dynamic striped phases play a role in YBa$_2$Cu$_3$O$_{6+x}$ superconductors.

Spin fluctuations are directly observable in neutron scattering experiments since the neutron has a magnetic moment that provides the interaction between the neutron and the unpaired electronic spins.  The spin fluctuations now have been quite extensively mapped out in the YBa$_2$Cu$_3$O$_{6+x}$ superconductors \cite{11,12} and while some features of the fluctuations appear to support the striped-phase picture, others do not.  In particular, for the highest $T_c$ materials there is little evidence that spin fluctuations occur with spatial distributions that support the simple stripe picture where antiphase spin boundaries occur at the position of charge stripes.  Nevertheless, the charge fluctuations themselves are interesting, and we concentrate on them in this paper.  Important questions are whether there is evidence for fluctuating charge stripes at optimum doping levels, and if these stripes exist, what is their temperature-dependence.

The basis for the investigation in this paper is the observation by triple-axis spectrometry that phonon branches acquire sizable anomalies at the momentum positions that are appropriate for the charge stripes of a stripe-phase \cite{10}.  The phonon investigated in [10] was the 
$\langle B_{2u}B_{3u}\rangle$ mode that results from oxygen motions in the Cu-O plane and is particularly sensitive to charge fluctuations.  Other phonons may also be affected by the charge fluctuations to some degree.  The ideal measurement would be to determine the phonon branches for a number of YBa$_2$Cu$_3$O$_{6+x}$ materials of different oxygen doping levels and to measure the temperature dependence of any anomalies found.  The low-temperature dispersion relation for the 
$\langle B_{2u}B_{3u}\rangle$
 planer oxygen mode has been determined by triple-axis spectrometry for the materials to be discussed in this paper \cite{13}.  However, these measurements take considerable spectrometer time and it would be impossible to make temperature-dependent measurements with the standard triple-axis technique.  An effective shortcut is the use the integration technique \cite{8,14}.  
This technique makes use of the fact that the scattering of interest 
can be added along the $c^\ast$ positions in the reciprocal lattice.

Fig. 1 shows the reciprocal lattice for YBa$_2$Cu$_3$O$_{6+x}$ for the $c^\ast$-$a^\ast(b^\ast)$ plane used in the measurement.   Since the crystals are twinned we cannot distinguish between the $a^\ast$ 
or $b^\ast$ directions.  The scattering of interest occurs on rods of scattering that 
extend along the $c^\ast$ axis as represented by the wavy lines in the figure.  In reality, 
the phonon cross section is not uniform along the wavy line because different positions on the 
line sample different displacement patterns.  This is not a problem as we are not necessarily trying to measure an individual phonon mode.  In the integration technique the spectrometer is set up so the incident neutron wave vector $k_i$ lands on one of the wavy lines and the outgoing wave vector 
$k_f$ extends along the wavy line.  No analyzer is used so that $k_f$ can be of any length, thus integrating the intensity along the line.  A scan (such as SCAN 1 in Fig. 1) is made by changing the spectrometer angles so that $k_f$ is moved along the direction of interest, in this case $a^\ast(b^\ast)$, while remaining parallel to the $c^\ast$ axis.  The technique is a powerful one and the incommensurate magnetic scattering in YBa$_2$Cu$_3$O$_{6+x}$ was originally found using this technique \cite{8}.

A dip or phonon anomaly will result in a peak in such a scan because of two effects.  The phonon cross section varies inversely as the phonon energy, so the lower phonon energy results in a larger scattering.  Secondly, the lower phonon energy means the energy transfer which is proportional to $k_i^2-k_f^2$ is smaller.  $k_i$ is kept fixed in length in the scan so that at lower phonon energies, $k_f$ is larger.  This means the counting rate is increased as the resolution volume sampled by the spectrometer is proportional to $k_f^3$.  Counting rates are high for this technique as many phonons are counted in the integration.  This results in a large background, but a signal of interest large enough to be observed on top of this background is detected.  The measured quantity is related to $S(q)$ which is given by the energy integral of the dynamic structure factor $S(q,\omega)$.

Fig. 2A and 2B show the results of measurements for YBa$_2$Cu$_3$O$_{6.45}$ (YBCO6.45) and 
YBa$_2$Cu$_3$O$_{6.6}$ (YBCO6.6).  The measurements were made on the HB-1 and HB-2 triple-axis spectrometers at the HFIR reactor at Oak Ridge National Laboratory. YBCO6.45 has a $T_c$ of 48K and elastic scattering shows charge ordering below 150 K at a wave vector of about 0.135 reciprocal lattice units (r.l.u.) along the [1, 0, 0] direction \cite{15}.  The integration technique will show a signal from both the charge ordering and charge fluctuations.  For this measurement a pyrolitic graphite monochromator was used with an incident neutron energy of 13.7 meV. Collimation was 48'-20'-20' from before the monochromator until after the sample. The integration scans made are represented by SCAN 1 in Fig. 1 and show peaks at about q = 0.135 r.l.u away from the (1, 0, 0) position in agreement with the elastic scattering measurements. Since the incident energy is 13.7 meV, the charge fluctuaations occur at rather low energies. The peaks are observed up to 300 K, which is the highest temperature of the measurement.  No broadening of the peaks is found so that the charge fluctuations are well defined at the highest temperature range of our measurement. 

For YBCO6.6 ($T_c = 62.5$ K), diffraction measurements show no static charge-order for temperatures as low as 10 K, and no charge fluctuations can be identified using an incident energy of 13.7 meV.  However, we know from triple-axis spectrometry that the 
$\langle B_{2u}B_{3u}\rangle$
 oxygen mode that occurs in the 40 meV energy range has a distinct phonon anomaly \cite{10}.  If we use the integration technique with an incident energy high enough to measure the 40 meV range, we also integrate over all the low-energy phonons.  We know from the measurements using an incident energy of 13.7 meV that there are no low-energy phonon anomalies visible, so that the low-energy phonons only contribute to the background.  The elastic spin and isotopic incoherent scattering also greatly increases the background.  We can remove the low-energy phonons and the elastic scattering by placing a pyrolitic graphite filter in the scattered beam.  The filter transmits essentially all the neutrons with energies less than 15 meV and removes higher energies to a greater or lesser extent dependent on the energy.  An ideal incident energy is 54.8 meV, as the filter in the scattered beam is highly efficient at removing this energy so that elastic scattering processes can be avoided.  In this case, since only neutrons with energies lower that 15 meV can be transmitted with very high-efficiency, the lowest energy sampled in the integration scan would be $54.8-15 = 39.8$ meV.  We need to include energies somewhat lower than this for YBCO6.6 as the phonon anomaly which occurs at about q = 0.2 r.l.u extends below this energy range \cite{10}.  We thus chose an incident neutron energy of 52.4 meV which covers the energy range of the phonon anomaly, but does not entirely exclude elastic scattering. In this case a Be monochromator was used with the (1, 0, 1) reflection in order to obatin better energy resolution. The spectrometer collimation was the same as for the 13.7 meV measurement.   The plots in Fig. 2B for YBCO6.6 show the results of the measurement.  The scan used is denoted SCAN 2 on Fig. 1 and the momentum transfer is referenced to the nearest Bragg reflection which is the (2, 0, 0).  The signal-to-noise is not particularly good as we cannot eliminate the elastic scattering, but a peak is observed at about 1.8 r.l.u. showing the phonon anomaly at q = 0.2 r.l.u as observed earlier \cite{10}.  Temperature-dependent measurements show that the peak seems to be present up to 300K.  The triple-axis measurements showed some broadening of the anomaly at 300K.  Because of the necessity to use a non-optimal incident energy, the integration measurements do not provide any improvement on the earlier triple-axis results.

We next extend our measurements to the compositions YBCO6.8 ($T_c = 82$ K) and YBCO6.95 ($T_c = 92.5$ K).  For these materials, low-temperature triple-axis measurements \cite{13} show that the phonon anomaly appears at slightly higher q-values where the phonon energy is higher.  We can now cover the energy of the phonon anomaly using an incident energy of 54.4 meV and make temperature dependent measurements.  For this energy the elastic scattering can be entirely eliminated by the filter and the integration technique becomes very effective.  The measurements for YBCO6.8 are shown in Fig. 3A.  The solid line is a fit to the data using a Gaussian distribution on a linearly sloping background.  Peaks are observed at q = $0.230\pm 0.007$ r.l.u. at 10 K and 150 K, while only a broad distribution is found at 200 K.  Thus, charge fluctuations are only observed below 200 K for YBCO6.8.  Fig. 3B shows the results for YBCO6.95.  Here peaks are observed at q = $0.235 \pm 0.0075$ r.l.u for 10 K and 75 K, but at 100 K the peak is greatly broadened.  At 150 K the peak is sufficiently broadened as to be invisible within the measurement error.  Therefore, the influence of charge fluctuations is only distinctly observed below about 100 K in YBCO6.95.

We have found that peaks in the integrated scattering from lattice vibrations can be identified at oxygen concentrations over the range from low to optimum doping.  We can directly connect the position of the peak for YBCO6.45 to the charge-ordering wavevector.  The peak for YBCO6.6 corresponds to the previously observed anomaly in the 
$\langle B_{2u}B_{3u}\rangle$
 phonon stemming from the motions of the planer oxygen atoms \cite{10}. 
Measurements made for YBCO6.8 and YBCO6.95 occur at positions of phonon anomalies observed at low temperatures by triple-axis spectrometry \cite{13}.  The present measurement confirms the triple-axis measurements and provides information at higher temperatures.  The wavevector of the phonon anomalies for the lower doping compounds is twice that of the lowest energy magnetic scattering.  This is consistent with the result expected for a striped phase with antiphase magnetic domain walls at the charge stripes.  We cannot make the same identification for YBCO6.8 and YBCO6.9, as there is no prominent magnetic scattering at one-half the wavevector of the observed phonon anomalies.  Thus, while our results are consistent with fluctuating charge-stripes there seems to be no matching set of magnetic stripes.  It appears that we can consider a picture for the highly-doped compositions in which there are fluctuating linear regions of charge, but the spin distribution shows no distinct disruption at the 
charge boundary.

The charge fluctuations become visible at high temperatures for the lightly doped compounds, and for YBCO6.45 occur at temperatures higher than room temperature.  The triple-axis measurements for YBCO6.6 showed some broadening at room temperature so that we can assume the charge fluctuations become visible in the neighborhood of 300 K.  For YBCO6.8 and YBCO6.93, charge fluctuations are found below 200 K and 150 K, respectively. 

Fig. 4 shows the results of the measurements in terms of the hole doping \cite{11} for the compositions used.  Fig. 4A shows that the wavevector of the phonon anomalies observed increases as the hole doping increases.  Associating the phonon anomalies with charge fluctuations would imply that the charge stripes become closer together as more charge is added.  Fig. 4B shows the temperatures where the phonon anomalies first appear.  The temperatures appear to be somewhat above the pseudogap \cite{16} temperature.  Unfortunately, the information we can obtain on the charge fluctuations is rather limited as we can only observe them indirectly through their effect on the atoms in the crystal lattice.  Nevertheless, the information we do have shows that a charge density wave or dynamic charge stripe 
exists in 
YBa$_2$Cu$_3$O$_{6+x}$
 for oxygen contents up to optimal doping.  The non-uniform charge distribution observed may play a role in the superconducting mechanism.

This work was supported by U.S. DOE under contract 
DE-AC05-00OR22725 with 
UT-Battelle, LLC.

\begin{figure}
\caption{
Reciprocal lattice of the scattering plane used in the experiment.  The measurement integrates along a path parallel to $c^\ast$ denoted by the wavy lines.  Scans were made at $(1 + q, 0, 0)$ and 
at $(2 - q, 0, 0)$ as shown by SCAN 1 and SCAN 2. 
}
\label{autonum}
\end{figure}

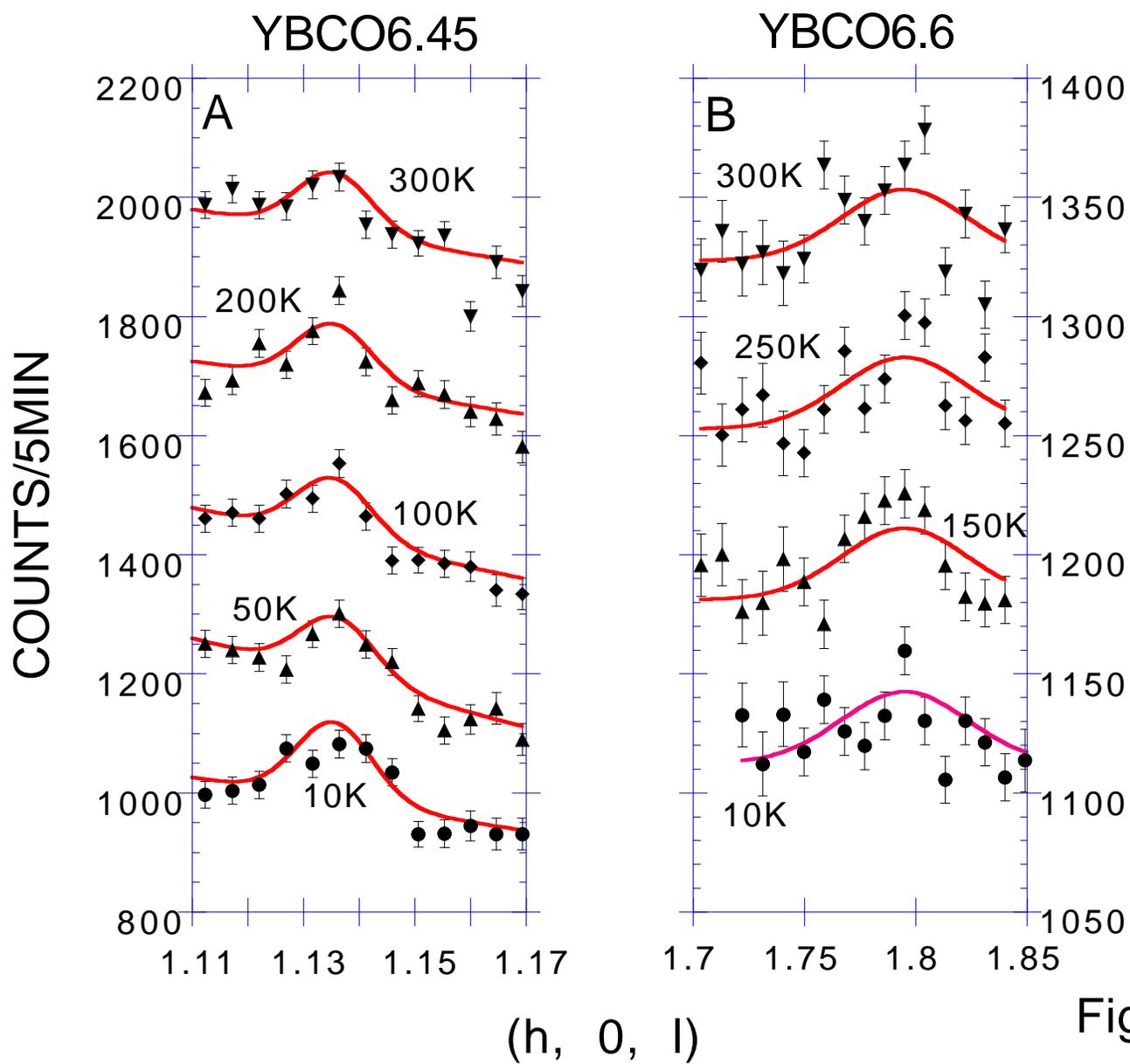
\begin{figure}
\caption{
A and B show integrating scan data for YBCO6.45 and YBCO6.6.  The incident neutron energy was 13.7 meV for YBCO6.45 and 52 meV for YBCO6.6.  A filter was used in the scattered beam to partially eliminate elastic and low-energy phonon background for YBCO6.6.  The solid lines are gaussian fits to the data on a linear background.  In A, 200 counts was added to the 50 K data, 400 to the 100 K, etc., so the scans could be compared on one graph.  In B, 75 counts were added at successive temperatures.  A number of runs were averaged to obtain the error bars shown. 
}
\end{figure}

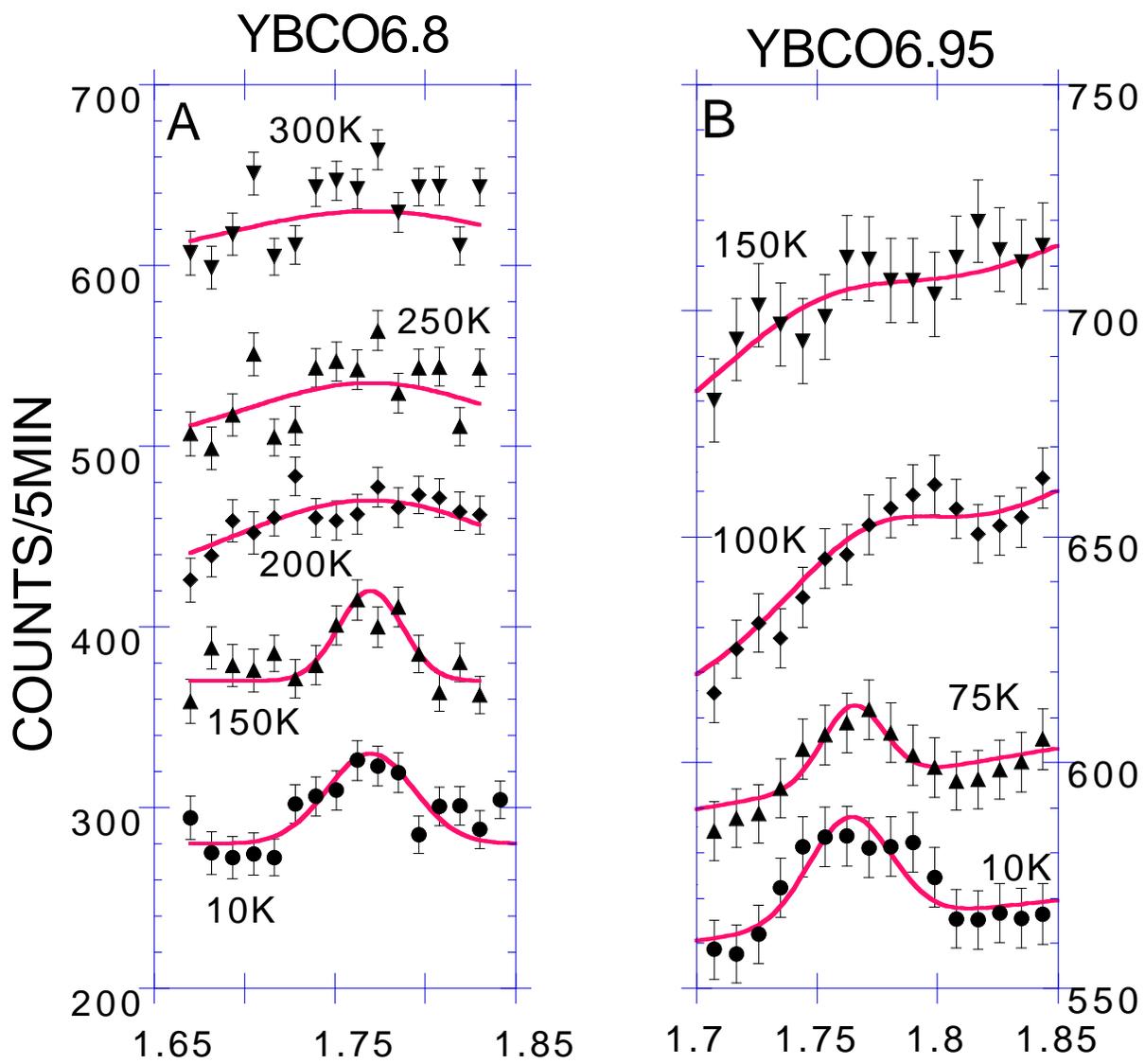
\begin{figure}
\caption{
A and B show integrating scan data for YBCO6.8 and YBCO6.95.  The incident neutron energy was 54.4 meV, and filters in the scattered beam largely eliminate elastic scattering and greatly reduce low-energy phonon scattering.  The solid lines are gaussian fits to the data on a linear background.  Fifty extra counts were added in A at each succeeding temperature so that the results could be compared on one graph.  The different signal-to-noise ratio for the two experiments reflects the fact that they were done on different spectrometers with different filters.  A number of runs were averaged together to obtain the error bars shown. 
}
\end{figure}

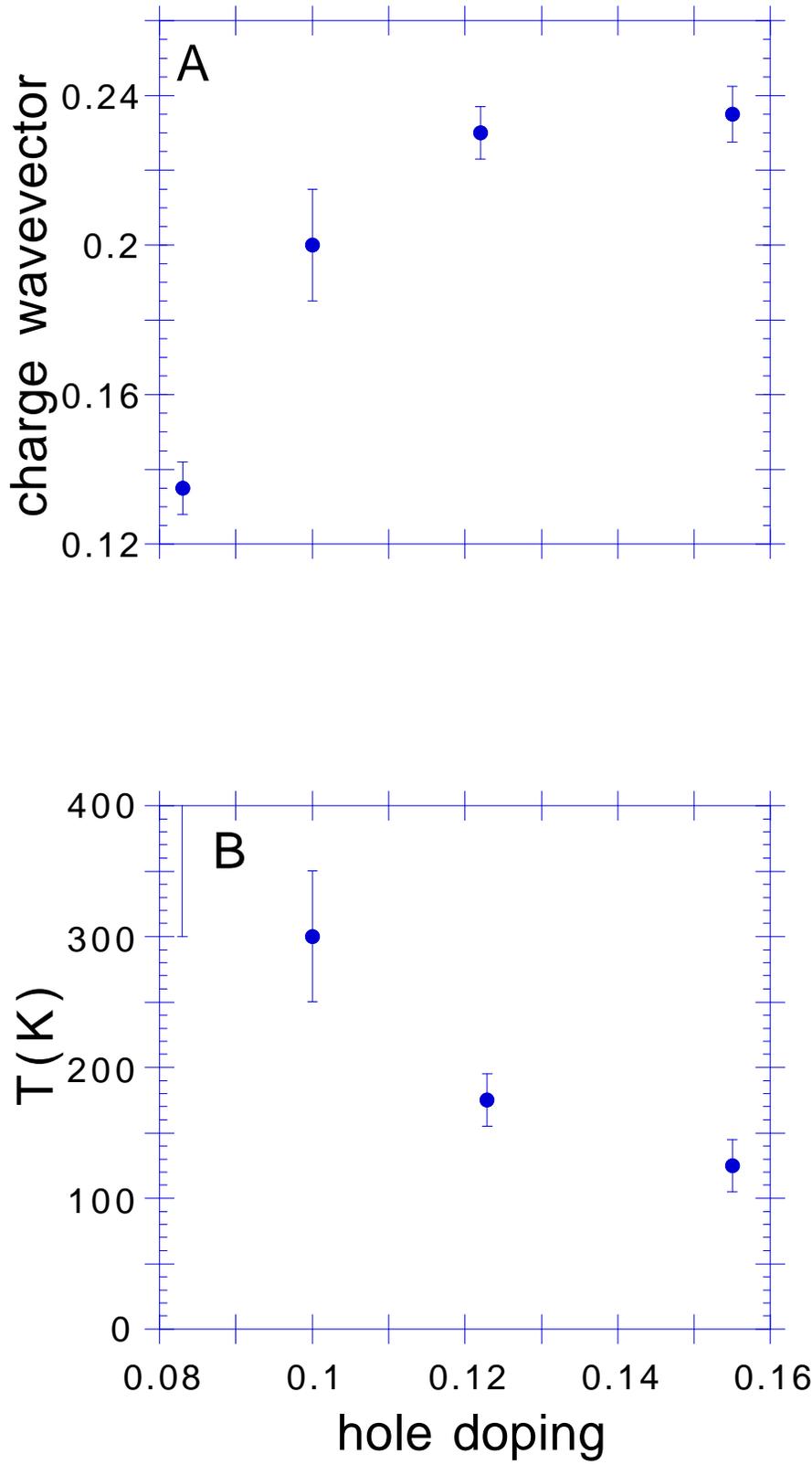
\begin{figure}
\caption{
A shows the wavevector of the phonon anomaly plotted vs. the hole content of the samples used in the experiment.  B shows the temperature at which the phonon anomaly is observed.  For the YBCO6.45 material (hole doping 0.05) we can only say that this temperature is above room temperature.  For YBCO6.6 (hole doping 0.1) the data point is taken from the earlier triple-axis data which shows substantial broadening of the anomaly at about room temperature.  For YBCO6.8 (hole doping 0.123) and YBCO6.95 (hole doping 0.155), the present measurement gives the onset temperature.          
}
\end{figure}
\end{document}